\documentclass[
nofootinbib,
 amsmath,amssymb,
 aps, physrev,
twocolumn
]{revtex4-1}

\usepackage[utf8]{inputenc}
\usepackage{graphicx}
\usepackage{dcolumn}
\usepackage{bm}

\usepackage[colorlinks,linkcolor=blue,citecolor=blue,urlcolor=blue]{hyperref}
\usepackage{bm,graphicx,dcolumn,epstopdf,epsf, latexsym,mathbbol, amssymb,amsmath,color,slashed, mathrsfs,mathcomp,simplewick}
\usepackage[center]{subfigure}
\usepackage{multirow}
\usepackage{graphicx}
\usepackage{makecell}
\usepackage{epstopdf}
\usepackage[table]{xcolor}
\usepackage{multirow}
\usepackage[table]{xcolor}
\usepackage{booktabs}
\usepackage{geometry}
\usepackage{orcidlink}
\usepackage{hyperref}
\usepackage{booktabs} 
\usepackage{comment}
\graphicspath{{figs/}} 

\begin{document}

\title{\textbf{Sterile neutrino dark matter from a TeV scale seesaw}}
\author{Kunal Pandey\orcidlink{0009-0006-0098-1517}}
\email{kpandey7007@gmail.com}
\author{Suhail Khan\orcidlink{0009-0007-4941-0069}}
\email{suhail187148@st.jmi.ac.in,suhail@ctp-jamia.res.in}
\affiliation{Centre for Theoretical Physics, Jamia Millia Islamia, New Delhi 110025, India}

\begin{abstract}

In this article, we investigate the phenomenological aspects of a feebly interacting sterile neutrino dark matter candidate within a low-scale seesaw framework. The Type-I seesaw model is augmented by a second complex scalar doublet ($\Phi_{\nu}$), which couples exclusively to the heavy right-handed neutrinos and the lepton doublet, thereby generating the neutrino Dirac mass term while the first scalar doublet is responsible for giving mass to the remaining Standard Model particles.
The lightest sterile neutrino ($N_1$) acts as a feebly interacting massive particle (FIMP), produced via decays of $W^\pm$, $Z$  and extra scalars present in the setup. We point out that $W^\pm$ and $Z$ contributions were overlooked in the previous studies, which actually dominate the $N_1$ production by a factor of $\sim 10^{13}$ and solely determines the relic abundance. Incorporating them leads to several novel consequences for the DM phenomenology like a new non-thermal condition which leads to smaller Yukawa couplings. We thoroughly discuss about the enhancement possibilities of $N_1$'s mass which is controlled by the small vacuum expectation value ($v_{\nu}$) of the second Higgs doublet. After incorporating the latest Lyman-$\alpha$ forest observations, this setup can accommodate both warm and cold dark matter scenarios.

\end{abstract}
\maketitle
\section{Introduction}

Among the numerous puzzling issues present in modern particle physics and cosmology are the origins of light neutrino masses~\cite{Fukuda:1998,Ahmad:2002,Ahn:2003} and the nature of dark matter (DM)~\cite{Bertone:2010at}. Numerous ideas have already been explored to address these problems, but probably one of the most intriguing questions is whether these two issues are correlated. At the very heart of the neutrino mass problem lies the practical issue of testability. It is now well established that the explanation of light neutrino masses introduces a very high scale in the theory ($\sim 10^{15}$ GeV), which is far beyond the current experimental reach. Hence, from a model-building point of view, it is desirable that the neutrino mass problem be probed by a lower-scale theory. To address these problems, one must go beyond the Standard Model (SM)~\cite{Feng:2010} and search for a minimal framework that could possibly link these two enigmas without introducing a high scale in the theory. In this spirit, the Type-I seesaw mechanism~\cite{Minkowski:1977sc,GellMann:1979vob,Mohapatra:1979ia,Schechter:1980gr,Schechter:1981cv} seems to be a promising framework, as it introduces three additional heavy right-handed neutrinos (RHNs), $\sim 10^{15}$ GeV, which are singlets under the SM gauge group. The newly introduced RHNs interact with the light neutrinos and the SM Higgs via Yukawa interactions. These Yukawa interactions not only control the scale of the light neutrino masses but also their interaction strength with the RHNs. It thus seems natural to study these Yukawa interactions and inquire whether they could also play a major role in addressing the DM problem. An elegant approach in this direction is to motivate one of the RHNs as the DM candidate. Numerous studies have been performed in this direction, for instance, in the original $\nu$MSM model \cite{Asaka2005b, Asaka2005a} the lightest RHN is the DM candidate with mass   $\sim$ keV, the Dodelson-Widrow (DW) mechanism~\cite{Dodelson:1993je} which utilizes the effective active-sterile mixing for the DM production, various U(1) gauge symmetric models \cite{Biswas2016,Okada2020,Eijima:2022dec,Seto:2024lik,Khalil:2008kp} which study the sterile neutrino production via decays of SM particles as well as newly introduced gauge bosons. The present study, addresses the phenomenological implications of identifying the lightest RHN ($N_1$) as the dark matter candidate with the aid of the neutrino-philic two-Higgs-doublet model ($\nu2$HDM)~\cite{Ma:2000cc}, which provides the low-scale seesaw framework. 
 
 In this framework, an extra complex scalar doublet ($\Phi_\nu$) in the $\nu2$HDM model with a small VEV ($v_{\nu} $ ) MeV is introduced, which is solely responsible for neutrino mass generation, while the SM Higgs doublet interacts in the usual manner. The small VEV opens up the possibility for the heavy RHN masses to attain lower values down to the TeV scale without requiring vanishingly small Yukawa couplings. However, it is not possible to arbitrarily reduce the scale of the RHNs, as then the Yukawa couplings must also assume lower values to induce suppressed light neutrino masses. Thus, if $N_1$ is to be motivated as a light dark matter candidate, the corresponding Dirac Yukawa couplings ($Y^{\nu}_{i1}$) must be significantly reduced to respect neutrino observables. This suggests that $N_1$ should be produced via out-of-equilibrium freeze-in production, which requires feeble couplings of dark matter with the thermal bath, rather than the freeze-out mechanism, which presumes that the dark matter was in thermal equilibrium with the rest of the particles. This class of particles for possible dark matter candidates is collectively known as Feebly Interacting Massive Particles (FIMPs)~\cite{Hall:2010}. Lots of explicit implementations of the FIMP framework have previously been studied \cite{McDonald:2001vt, Asaka:2005cn, Asaka:2006fs,Kusenko:2006rh, Cheung:2011nn, Yaguna:2011qn, Yaguna:2011ei,Blennow:2013jba, Chu:2013jja, Merle:2013wta,Klasen:2013ypa}, and with a singlet scalar \cite{McDonald:2001vt,Yaguna:2011qn,Yaguna:2011ei} or a new singlet fermion \cite{Klasen:2013ypa} are the two simplest extensions of the SM that incorporate a FIMP because FIMPs are by definition singlets under the Standard Model (SM) gauge group. Under the freeze-in scenario, $N_1$ gets almost decoupled from the seesaw sector, and the remaining two RHNs play the dominant role in neutrino mass generation and could also participate in the creation of the matter-antimatter asymmetry.
 
  The production of $N_1$ proceeds via two-body decay processes of the extra scalars in the setup and also from the decays of the SM gauge bosons ($W$ and $Z$). This possibility was partially introduced in \cite{Adulpravitchai:2015mna} but not thoroughly investigated.
  The novelty of this article lies in the fact that, we have shown that, the previous study of sterile neutrino production in the neutrino-philic two-Higgs-doublet model ($\nu2$HDM)~\cite{Adulpravitchai:2015mna} has ignored the most dominant sterile neutrino production channels ($W$ and $Z$ decays) which arise due to the consideration of heavy-light mixing. Realization of this fact leads to many non-trivial results in the DM phenomenology, for instance, in a revised non-thermal condition which points towards smaller values of the Dirac Yukawa couplings ($Y^{\nu}_{i1}$). Another interesting outcome of the analysis is that the dark matter relic abundance depends on: the Dirac Yukawa couplings ($Y^{\nu}_{i1}$), DM mass ($M_{N_1}$) and vev of the second doublet ($v_{\nu}$). In fact, by choosing an appropriate value of $v_{\nu}$ the parameter space for allowed  DM mass ($M_{N_1}$) could be enhanced (even upto GeV scale). Also, for a give value of $v_{\nu}$, the lightest neutrino mass scale ($m_{\nu_1}$) is uniquely fixed by the DM requirements. As a result, an interesting correlation is obtained between the seesaw scale, dark matter, and the lightest neutrino mass. Considering the latest bounds coming from Lyman-$\alpha$ results, it is also found that the setup naturally accommodates both warm and cold dark matter scenarios and provides a natural explanation for the much-debated $3.55$ keV X-ray line.

Our work is organized as follows. In Section~\ref{model}, we discuss our model in the context of the nature of dark matter and the non- thermal condition for freeze-in mechanism. In Section~\ref{sec:Production}, we carefully study the production of FIMP dark matter from the decay of gauge bosons; this channel dominates all other production channels by a factor of $10^{13}$, allowing us to obtain the corresponding enhanced viable parameter space for the relic density. Section~\ref{sec:constraint} is dedicated to the discussion on constraints arising from small-scale structure formation. Finally, we present our conclusions in Section~\ref{sec:con}.

\section{The Model}\label{model}

Originally, the neutrino-philic two Higgs doublet model ($\nu$2HDM) was proposed in~\cite{Ma:2000cc}. In this model, the scalar sector of the Standard Model (SM) is extended by an additional doublet (the neutrinophilic doublet) $\Phi_\nu$ with the same quantum numbers as the SM Higgs doublet $\Phi$. Three heavy right-handed neutrinos ($N_{R_i}$) are also added in accordance with the standard Type-I seesaw mechanism to realize light neutrino masses. However, the standard Type-I seesaw interaction $L_i \tilde{\Phi} N_{R_i}$ is forbidden by imposing a global $U(1)_L$ symmetry under which $L_\Phi = 0$, $L_{\Phi_\nu} = -1$, and $L_{N_{R_i}} = 0$. This ensures that $\Phi_\nu$ couples exclusively to $N$, while $\Phi$ couples to quarks and charged leptons as in the SM and is responsible for their masses.

The two scalar doublets as follows:
\begin{equation}
\displaystyle
\Phi = 
\begin{pmatrix}
\phi^+ \\
\frac{v + \phi^0_r + i \phi^0_i}{\sqrt{2}}
\end{pmatrix}, \quad
\Phi_\nu =
\begin{pmatrix}
\phi^+_\nu \\
\frac{v_\nu + \phi^0_{r\nu} + i \phi^0_{i\nu}}{\sqrt{2}}
\end{pmatrix}.
\label{doublet}
\end{equation}

The corresponding Higgs potential is then
\begin{align}
V(\Phi,\Phi_\nu) ={}& m^2_\Phi\,\Phi^\dagger\Phi + m^2_{\Phi_\nu}\,\Phi^\dagger_\nu\Phi_\nu
+ \frac{\lambda_1}{2} (\Phi^\dagger\Phi)^2 
+ \frac{\lambda_2}{2} (\Phi^\dagger_\nu\Phi_\nu)^2 \nonumber \\
& + \lambda_3 (\Phi^\dagger\Phi)(\Phi^\dagger_\nu\Phi_\nu) 
+ \lambda_4 (\Phi^\dagger\Phi_\nu)(\Phi^\dagger_\nu\Phi) \nonumber \\
& - (\mu^2 \Phi^\dagger\Phi_\nu + \text{h.c.}).
\label{pot}
\end{align}

where a soft symmetry breaking term $\mu^2$ has also been considered which breaks the $U(1)_L$ symmetry explicitly but softly. In order to calculate the VEV's of the two Higgs doublets in terms of parameters of the scalar potential, thus the following minimization condition:
\begin{align}
v\, \left[ m^2_\Phi + \frac{\lambda_1}{2} v^2 + \frac{\lambda_3 + \lambda_4}{2} v^2_\nu \right] - \mu^2 v_\nu = 0 
 \\
v_\nu\, \left[ m^2_{\Phi_\nu} + \frac{\lambda_2}{2} v^2_\nu + \frac{\lambda_3 + \lambda_4}{2} v^2 \right] - \mu^2 v = 0.
\end{align}

Considering the following choices for the parameters:
\begin{equation}
m^2_\Phi < 0,\quad m^2_{\Phi_\nu} > 0, \quad |\mu^2| \ll m^2_{\Phi_\nu},
\end{equation}

the two VEV's are derived as follows:
\begin{equation}\label{VEVs}
v \simeq \sqrt{-2 m^2_\Phi/\lambda_1}, \qquad v_\nu \simeq \frac{\mu^2 v}{m^2_{\Phi_\nu} + (\lambda_3 + \lambda_4)v^2/2}.
\end{equation}

From the Eq.~(\ref{VEVs}), it follows that $v_\nu \sim 1\,\text{GeV}$ is obtained with $\mu \sim 10\,\text{GeV}$ and $M_{\Phi_\nu} \sim 100\,\text{GeV}$. An important point to note here is that the only source of $U(1)_L$ breaking is the $\mu^2$ term, so the radiative corrections to $\mu^2$ are proportional to $\mu^2$ itself and will only be logarithmically sensitive to the cutoff~\cite{Davidson:2009ha}. This ensures that the obtained VEV hierarchy $v_\nu \ll v$ is stable against radiative corrections~\cite{Haba2011, Morozumi2014, Morozumi2012}. After spontaneous symmetry breaking (SSB), the physical Higgs bosons are obtained, which contain admixtures of the components of both doublets and are given by~\cite{Guo2017}:

\begin{align}
S^{+} = \phi^+_\nu\,\cos\beta - \phi^+\,\sin\beta, \quad A^{0} = \phi^0_{i\nu}\,\cos\beta - \phi^0_i\,\sin\beta,
\\ 
S^{0} = \phi^0_{r\nu}\,\cos\alpha - \phi^0_r\,\sin\alpha, \quad h = \phi^0_r\,\cos\alpha + \phi^0_{r\nu}\,\sin\alpha,
\end{align}
and the mixing angles $\beta$ and $\alpha$ are given as:
\begin{equation}
\tan\beta = \frac{v_\nu}{v}, \qquad 
\tan 2\alpha \simeq \frac{2 v_\nu v\, [-\mu^2 + (\lambda_3 + \lambda_4) v v_\nu]}{-\mu^2 + \lambda_1 v v_\nu}.
\end{equation}

After neglecting the terms of $\mathcal{O}(v_\nu^2)$ and $\mathcal{O}(\mu^2)$, masses of the bosons are given as follows:
\begin{align}
m^2_{S^+} &\simeq m^2_{\Phi_\nu} + \frac{1}{2}\lambda_3 v^2, \nonumber \\
m^2_{A^{0}} &\simeq m^2_{S^{0}} \simeq m^2_{S^+} + \frac{1}{2}\lambda_4 v^2, \nonumber \\
m^2_h &\simeq \lambda_1 v^2.
\end{align}

It is to be noted that the mixing angles are suppressed by the small value of $v_\nu$; hence, $h$ is almost identically the $ 125$ GeV SM Higgs boson~\cite{Chatrchyan2012,Aad2012}. For simplicity, we adopt a degenerate mass spectrum of $\Phi_\nu$ as $m_{s^+} \simeq m_{S^0} \simeq m_{A^0} \simeq $ $m_{\Phi_\nu} $ in our analysis, which are allowed by various constraints~\cite{Machado2015}.

The relevant parts of the Yukawa interaction and the mass term for the heavy right-handed neutrinos are given as:

\begin{equation}\label{yukint}
-\mathcal{L}_Y \supset Y^{\nu}_{ij} \Bar{L_{i}}\tilde{\Phi}_\nu N_{R_{j}} +  \frac{1}{2} \Bar{N_{R_{i}}^c} M_{R_{i}} N_{R_{i}} + \text{h.c.}
\end{equation}
where $\tilde{\Phi}_\nu = i \sigma_2 \Phi_\nu^*$, $L$ is the lepton doublet, and the second term forms the mass matrix of the heavy right-handed neutrinos, which is assumed to be diagonal for simplicity. After spontaneous symmetry breaking (SSB), the above interaction Lagrangian is written as:

\begin{equation}
-\mathcal{L}_Y \supset \frac{1}{2} 
\begin{pmatrix}
\bar{\nu_L^c} & \bar{N_R}
\end{pmatrix}
\mathcal{M}
\begin{pmatrix}
\nu_L \\
N^c_R
\end{pmatrix}
+ \text{H.c.},
\label{eq:Lnu}
\end{equation}
where $\mathcal{M}$ is the $6 \times 6$ mass matrix and is written as:
\begin{equation}
\mathcal{M} =
\begin{pmatrix}
0 & M_D \\
M_D^T & M_R
\end{pmatrix},
\label{eq:massmatrix}
\end{equation}
$M_D$, $M_R$ are $3 \times 3$ neutrino Dirac mass matrix and heavy Majorana neutrino mass matrix, respectively. In order to get the eigenvalues, $\mathcal{M}$ can be diagonalized using a unitary matrix $\mathcal{U}$ as follows:

\begin{equation}
\mathcal{M}_{\text{diag}} = \mathcal{U}^T \mathcal{M} \mathcal{U},
\label{eq:diag}
\end{equation}
thereby giving the masses of the neutrinos in the physical basis. The mass eigenstates can be defined via the unitary rotation:
\begin{equation}
\begin{pmatrix}
\nu_L \\
N^c_R
\end{pmatrix}
= \mathcal{U}
\begin{pmatrix}
\nu_i \\
N_{Rj}
\end{pmatrix},
\label{eq:rotation}
\end{equation}
and the matrix $\mathcal{U}$ can be expressed as (expanding in terms of $M_D M_R^{-1}$)
\begin{equation}
\mathcal{U} =
\begin{pmatrix}
U_{\nu\nu} & U_{\nu N} \\
U_{N\nu} & U_{NN}
\end{pmatrix}.
\label{eq:Ublocks}
\end{equation}
To leading order, one finds~\cite{Casas:2001sr, Ibarra:2003up}
\begin{align}\nonumber
U_{\nu\nu} &\simeq U_{\text{PMNS}}, \\ \nonumber
U_{\nu N} &\simeq M_D^\dagger M_R^{-1}, \nonumber \\
U_{N\nu} &\simeq -M_R^{-1} M_D U_{\nu\nu}, \nonumber \\  
U_{NN} &\simeq I.
\end{align}
Here, $U_{\rm PMNS}$ is the PMNS (Pontecorvo-Maki-Nakagawa-Sakata) matrix~\cite{Fogli:2006,Zyla:2020,Esteban:2020}. By using Eq.~(\ref{eq:diag}) and Eq.~(\ref{eq:Ublocks}), one can quantify the mixing between different components of the seesaw. The light neutrino mass matrix is given by the seesaw formula~\cite{Minkowski:1977sc} as:
\begin{equation}
m_\nu = -\frac{v_\nu^2}{2} y M_R^{-1} y^T = U_{\rm PMNS}\, \hat{m}_\nu \, U_{\rm PMNS}^T,
\end{equation}
where $\hat{m}_\nu = \mathrm{diag}(m_1, m_2, m_3)$ is the diagonalized neutrino mass matrix. It is evident that due to the smallness of $v_\nu$, a TeV scale of $M_{N_{i}}$ could lead to light neutrino masses at the $0.1\,\mathrm{eV}$ scale.

\subsection{Sterile Neutrino Dark Matter}
One of the heavy right-handed neutrinos $N_1$ (dropped $R$ index) is motivated to be the dark matter candidate by keeping its mass much less than the masses of the $W$, $Z$, and the scalars present in the model. 
This allows for its production via heavy-light mixing (for $W$ and $Z$ bosons) and also via the Yukawa interaction. The masses of the remaining heavy neutrinos are assumed to lie far above the electroweak scale to satisfy neutrino observables.

A major requirement for the production of a FIMP dark matter~\cite{Hall:2009bx} is that its interactions with other particles in the thermal bath are feeble. This implies that the dark matter was never in equilibrium with the thermal plasma throughout its thermal evolution, necessitating very small Yukawa couplings of $N_1$ with the lepton doublet and the second Higgs doublet ($\Phi_\nu$).

As a result, $N_1$ effectively gets completely decoupled and does not participate in neutrino mass generation, leading to only two massive light neutrinos. For calculating light neutrino masses and mixing, the Casas-Ibarra parametrization~\cite{Casas:2001sr} can be employed, which expresses the Yukawa couplings in terms of experimental data on neutrino masses and mixing angles. Hence, for the Yukawa matrix $Y^{\nu}$ entries, one can write:

\begin{equation}
M_D = -i U_{PMNS} D_{\sqrt{m}} R^T D_{\sqrt{M}},
\label{pmns}
\end{equation}
where $D_m$ is the diagonal active neutrino mass matrix, $D_M$ is the diagonal heavy right-handed neutrino mass matrix, and $R$ is a complex orthogonal matrix which, for the present scenario, can be chosen to be of the following form:

\begin{equation}
R =
\begin{pmatrix}
1 & 0 & 0 \\
0 & \cos\theta_R & \sin\theta_R \\
0 & -\sin\theta_R & \cos\theta_R
\end{pmatrix}.
\label{matix}
\end{equation}
where, $\theta_R$ is generally complex. This description guarantees that the setup remains compatible with current neutrino experimental observables of mass-squared differences and all mixing angles. With this choice, $N_1$ is completely segregated from the rest of the particles in the setup and cannot be produced by any interactions. However, as discussed earlier, feeble couplings for $N_1$ are desired, and this can be circumvented by perturbing ($ Y_{i1}^{\nu} \sim\epsilon_{i1} <<1$) the matrix $M_D$, resulting in very small but non-zero entries in the first column. This leads to very small couplings of $N_1$ with the bath particles, which, if kinematically allowed, will enable production via decay of bath particles.

The tininess of the couplings depends on the stability of the dark matter candidate $N_1$ and is controlled by relic density requirements. It also dictates the scale of the lightest neutrino mass in this setup. A crucial point not explored in previous studies~\cite{Adulpravitchai:2015mna} is that the Yukawa interactions of these heavy right-handed neutrinos also enforce their interactions with Standard Model gauge bosons through heavy-light mixing with active neutrinos. The ignorance of these interactions is often justified due to the large masses of heavy right-handed neutrinos, rendering the heavy-light mixing $V_{ij} \sim M_{D_{ij}}/M_{N_{j}}$ extremely small to play any significant role in dark matter requirements. However, as will be shown, under the FIMP scenario, such small values of $V_{ij}$ are valuable and provide an alternative production channel for the dark matter candidate.

Thus, the interactions~\cite{Barman2023} of the heavy right-handed neutrinos with light neutrinos and the Standard Model gauge bosons are presented below, utilizing the heavy-light mixing:

\begin{equation}\label{gauge}
    \begin{aligned}
    - \mathcal{L}_{g} \subset & \frac{g}{\sqrt{2}} W_{\mu}^{+} \sum_{i, j=1}^3 [\Bar{N}_{i}^c (V^{\dagger})_{ij} \gamma^{\mu} P_{L} l_{j} ]  +  \frac{g}{2 C_{\theta_{w}}} Z_{\mu} \\
 \times   &  \sum_{i, j=1}^3 [\Bar{\nu_{i}} (U^{\dagger}_{PMNS} V)_{ij} \gamma^{\mu} P_{L} N_{{j}}^c + \Bar{N_{{i}}}^c (V^{\dagger}V)_{ij} \gamma^{\mu} P_{L} N_{{j}}^c ]
   \end{aligned}
\end{equation}
where $P_{L} = \frac{1 - \gamma^5}{2}$, $g$ is the gauge coupling constant, $C_{\theta_{w}} = \cos\theta_{w}$, $\theta_{w}$ is the weak mixing angle, and $V_{ij} = \frac{M_{D_{ij}}}{M_{N_{j}}}$ is the heavy-light mixing matrix element through which the heavy right-handed neutrinos interact with the $W$ and $Z$ bosons. This very fact was completely overlooked in the previous study, and it turns out that the actual dominant production of $N_1$ will proceed via $W^\pm$ and $Z$ boson two-body decays as per the above interaction Lagrangian ~(\ref{gauge}).

It should be noted that active neutrino observables are accounted by the remaining two right-handed neutrinos, all at the TeV scale. The studies regarding the neutrino sector of the model and also the explanation of the baryonic asymmetry are well known and therefore, that analysis is not repeated here and the focus is placed on the study of dark matter phenomenology.

\subsection{Non-thermal Regime}
\label{sec:out}

One of the major requirements of the freeze-in mechanism (FIMP) is that the dark matter particle should not reach thermal equilibrium with the rest of the particles present in the thermal bath. In the setup under consideration, only the gauge singlet heavy right-handed neutrinos could play the role of FIMPs. The thermal equilibrium in the early universe can be prevented only if their Yukawa interactions are sufficiently suppressed. In this section, the different processes that can produce these singlets are analyzed, and the necessary conditions for the out-of-equilibrium requirement are determined.  It is shown that only one of these, chosen to be $N_1$, can play the role of a FIMP. 

All the right handed neutrinos have the Yukawa interactions of the form $\Bar{L_{i}}\tilde{\Phi}_\nu N_{R_{j}}$ so, they can be produced via the two-body decay of the scalars present in the model and the decay width is approximately given by (see Eqs.~(\ref{GS1}) to (\ref{GS2}))
\[
\Gamma(\text{Scalars}\to N_1\,L) \sim M_{\text{Scalars}}\,|Y_{i1}^{\nu}|^{2} .
\]
Here $M_{\text{Scalars}}$ is the mass of scalars. However, apart from the scalar decay modes, two crucial decay channels involving W and Z bosons also exist, which utilize the mixing between $N_1$ and the light neutrinos. In presence of several production modes, the out-of-equilibrium condition is  determined by the fastest decay mode. It turns out that the fastest one out of all these available decay channels is that of Z boson whose decay width is given as: 
\[
\Gamma_{Z \to N \nu_i + N \bar{\nu}_i} = \frac{1}{48\pi} m_Z |Y^{\nu}_{i1}|^2 f(M_{N_1}^2 / m_Z^2), 
\]
where $f(a) = (1 - a)^2 (1 + 2/a)$. Then the out of equilibrium condition is given as:
\begin{eqnarray}
    \Gamma(Z \to N_1 \nu_i + N_1 \bar{\nu}_i)\lesssim H(T\sim M_{Z})\,,
\end{eqnarray}

where $H(T\sim M_Z)$ is the Hubble parameter ($H(T)= 1.66 \sqrt{g_{*}(T)} T^2/M_{P}$ and $M_{P}= 1.22 \times 10^{19} \text{GeV}$) evaluated at the mass of $Z$ boson $(M_Z)$. From the above inequality one can derive the following condition on the Yukawa couplings of $N_1$ as:

\begin{equation}\label{ybound}
\sum_{i=e,\mu,\tau} |Y_{i1}^{\nu}|^2 \lesssim 1 \times 10^{-18} \left( \frac{M_{N_{1}}}{\mathrm{GeV}} \right)^2, 
\end{equation}

So, for a given value of $M_{N_1}$, above inequality will determine the allowed values of $|Y_{i1}|$ and one is bound to respect this condition in order to make sure that the DM never thermalizes which is the main assumption of the freeze-in scenario. However, if only the scalar contributions are considered as in the previous study \cite{Adulpravitchai:2015mna}, then the resulting out-of-equilibrium condition looks like: 
\begin{eqnarray}
    \Gamma(\text{Scalars}\to N_1\,\Bar{L})\lesssim H(T\sim M_{\text{Scalar}})\,,
\end{eqnarray}

the corresponding condition on the Yukawa couplings of $N_1$ are obtained to be:
\begin{equation}
    \sum_{i=e,\mu,\tau} |Y_{i1}^{\nu}|^2 \lesssim 7 \times 10^{-17} \left(\frac{M_{\text{scalar}}}{ \text{GeV}}\right)
\end{equation}

Clearly, consideration of SM gauge boson decays result in a very tight constraint on the allowed couplings of the DM. For instance, if one chooses $M_{N_1} \sim 1\, \text{keV}$ then one has to obey: $  \sum_{i} |Y_{i1}^{\nu}|\lesssim 10^{-15}$  and for $M_{N_1} \sim 0.1\, \text{GeV}$ one has the condition: $  \sum_{i} |Y_{i1}^{\nu}|\lesssim 10^{-10}$. Contrastingly, if one (wrongly) considers only the scalar contribution (Eq.(25)) then the non-thermal condition depends only on the scalar mass. So, if one chooses $M_{\text{scalar}}$\,$ (200\, \text{GeV} -  1\, \text{TeV})$ then it gives: $  \sum_{i} |Y_{i1}^{\nu}|\lesssim (10^{-7}-2.6 \times 10^{-7})$.

Hence, we conclude that the if the gauge boson channels are excluded then this will drastically change the crucial non-thermal condition which, as a result, steer us towards incorrect values of the allowed couplings of DM with the thermal bath.
 
Following Eq.(23), it is clear that one has to have tiny Yukawa couplings for $N_1$ to be a DM candidate. This results in small values in the first column of $M_D$ implying that $N_1$ plays no role in the mass generation of active neutrinos. This conclusion is in complete accordance with that of \cite{Hernandez2014b,Hernandez2014a}, where the authors have argued that in low-scale Type-I seesaw models with three heavy neutrinos, if the mass of the lightest active neutrino $< 10^{-3}$ eV, then only one of the heavy sterile states never thermalizes. However, as discussed in subsequent sections, the requirement of the correct relic density of dark matter will make the mass of the lightest neutrino attain even further smaller values. It must be pointed out that if $N_1$ was never in thermal equilibrium, then the only mechanism of its production in the early universe is via freeze-in. However, $N_1$ can also be produced via the decay of the heavier singlets ($N_{2,3}$) or via $2\to 2$ scatterings of SM leptons or the scalars in the setup. However, all these processes are rather subdominant and do not modify the above equilibrium conditions. 

Models of sterile neutrino ought to have the active-sterile mixing (or, this is just the heavy-light mixing as in Type-I seesaw) and then, apart from the production via decay of heavy particles (freeze-in scenario), which we are considering in this work, another mechanism also exists, which also utilizes this heavy-light mixing. 

Sterile neutrinos are also produced through non-resonant oscillations of active neutrinos (see ref. Dodelson-Widrow \cite{Dodelson:1993je}). As a result, a primordial population of sterile neutrinos is obtained with a warm dark matter spectrum, unlike the usual cold dark matter candidates. The contribution to the relic abundance of sterile neutrino is given as \cite{Abazajian2001}:

\begin{eqnarray}
    \Omega_{\text{DW}} h^2 \approx 0.3 \times \left( \frac{\sin^2 2\theta}{10^{-10}} \right) \left( \frac{M_{N_1}}{100 \text{ keV}} \right)^2
\end{eqnarray}

where $\theta$ is the active-sterile mixing. However, this mechanism of sterile neutrino production is strongly constrained by Lyman-$\alpha$ bounds \cite{Boyarsky2009} and also X-ray observations \cite{Essig2013}, and it is now well accepted \cite{Boyarsky2009b,Seljak2006} that this production pathway cannot account for all the observable dark matter in the Universe. But, due to the presence of active-sterile mixing, this contribution to the relic density always exists and must be taken into account. For our case, we have found that for the $N_1$ mass range $1$ keV to GeV scale and the Yukawa couplings of the order $\sim 10^{-14}$, the contribution is quite less and hence irrelevant in our study.

So, the conclusion is that in order to simultaneously account for light neutrino masses and ask for a TeV scale of seesaw, only one FIMP candidate is allowed. There are four free parameters:\,  $m_{\phi_\nu}$,$M_{N_1}$, $Y^{\nu}_{i1}$, and $v_{\nu}$ which will participate  in the dark matter phenomenology. The rationale for treating $m_{\phi_\nu}$ as a free parameter is that its specific value does not influence the non-thermal freeze-in production or the dark matter phenomenology central to this study. Consequently, selecting any mass, like 200 GeV, for illustrative purposes does not affect our results. After fixing it, now there are three free parameters in our analysis.

In the upcoming section, various production channels will be discussed in detail, and it will be demonstrated that the observed dark matter relic density can be easily accounted for in this framework.

\section{Dark Matter Production}
\label{sec:Production}
This section discusses the various available production channels for the dark matter candidate $N_1$. Since $N_1$ has a direct coupling to the lepton doublet and to the neutrinophilic doublet, the relevant production pathway is by the decays of the scalars ($S^0,A^0,S^\pm,h$), which are in equilibrium with the thermal bath. The SM gauge bosons $W$ and $Z$ will also take part in the production mechanism with the aid of heavy-light mixing. The $N_1$ yield, $Y_{N_{1}}(T)=n_{N_1}(T)/s(T)$, is computed by solving 
the following Boltzmann equation \cite{Biswas2016}

\begin{align}
\frac{dY_{N_1}(z)}{dz} &= \frac{2 M_{pl} z}{1.66 m_h^2} \frac{\sqrt{g_*(z)}}{g_s(z)}    
\Big[ Y_Z^{eq} \big\langle\Gamma_{Z\to N_1 \nu}\big\rangle + Y_{A^{0}}^{eq} \big\langle\Gamma_{A^{0}\to N_1 \nu}\big\rangle \nonumber \\ 
& + Y_h^{eq} \big\langle\Gamma_{h\to N_1 \nu}\big\rangle \nonumber + Y_{S^{0}}^{eq} \big\langle\Gamma_{S^{0}\to N_1 \nu}\big\rangle 
\nonumber \\
& + Y_{S^{\pm}}^{eq} \big\langle\Gamma_{S^{\pm}\to N_1 \nu}\big\rangle + Y_W^{eq} \big\langle\Gamma_{W^{\pm}\to N_1\ell^{\pm}}\big\rangle\Big],\label{1BE}
\end{align}

with \(z = m_h/T\), the functions \(g_\rho(z)\) and \(g_s(z)\) are the effective degrees of freedom for the energy density (\(\rho\)) and entropy density (\(s\)) of the universe, respectively, their values are 110.75. The quantity \(\langle \Gamma_{A\to BC}\rangle\) represents the thermally averaged decay width~\cite{Babu2014} and is defined as:

\begin{eqnarray}
    \big\langle \Gamma_{A\to BC}\big\rangle = \frac{K_1(z)}{K_2(z)} \Gamma_{A\to BC}
\end{eqnarray}

where $K_1(z)$, and $K_2(z)$ are the modified Bessel functions of order $1$ and $2$, respectively.

The decay rates that enter into this expression are given as ~\cite{Coy:2021sse,Adulpravitchai:2015mna}:  
\begin{align}\label{GS1}
	\Gamma\left(S^{0}\to N_{1} \, \nu_{i}  \right) & =  \frac{m_{S^{0}}\,\left| Y^{\nu}_{i 1} \right|^{2}}{32\,\pi}\left(1-\frac{M_{N_1}^{2}}{m_{S^{0}}^{2}}\right)^{2}\ \nonumber,\\& \approx \frac{m_{S^{0}}\,\left| Y^{\nu}_{i 1} \right|^{2}}{32\,\pi},\nonumber \\
\end{align}

\begin{align}   
\Gamma\left(A^{0}	\to N_{1} \, \nu_{i}  \right) & =  \frac{m_{A^{0}}\,\left| Y^{\nu}_{i 1} \right|^{2}}{32\,\pi}\left(1-\frac{M_{N_1}^{2}}{m_{A^{0}}^{2}}\right)^{2}\ \nonumber,\\& \approx \frac{m_{A^{0}}\,\left| Y^{\nu}_{i 1} \right|^{2}}{32\,\pi},\nonumber \\
\end{align}

\begin{align}\label{GS2}
    \Gamma\left(S^{\pm}	\to N_{1} \, \overline{\ell_{i}^\pm}  \right) & =  \frac{m_{S^{\pm}}\,\left| Y^{\nu}_{i 1} \right|^{2}}{16\,\pi}\left(1-\frac{M_{N_1}^{2}}{m_{S^{\pm}}^{2}}\right)^{2}\ \nonumber,\\& \approx \frac{m_{S^{\pm}}\,\left| Y^{\nu}_{i 1} \right|^{2}}{16\,\pi} \nonumber.\\
\end{align}   

\begin{align}\label{hdecay}
    \Gamma\left(h	\to N_{1} \, \nu_i  \right) & =  \frac{m_{h} \,\left| Y^{\nu}_{i 1} \right|^{2}}{16\,\pi}\left(1-\frac{M_{N_1}^{2}}{m_{h}^{2}}\right)^{2}\ \nonumber,\\& \approx \frac{m_{h}\,\left| Y^{\nu}_{i 1} \right|^{2}}{16\,\pi} \nonumber.\\
\end{align}   

\begin{eqnarray} \label{wdecay}   
    \Gamma(W^{\pm} \to N_1 \ell^{\pm}_i) & =& \frac{1}{48\pi} m_W |Y_{i 1}^\nu|^2 f(M_{N_1}^2 / m_W^2), \\
     \Gamma(Z \to N_1 \nu_i + N_1 \bar{\nu}_i) &=& \frac{1}{48\pi} m_Z |Y_{i 1}^\nu|^2 f(M_{N_1}^2 / m_Z^2), \label{zdecay}   
\end{eqnarray}
where\quad
$
f(a) = (1 - a)^2 (1 + 2/a).
$
The approximations used in Eq.~(\ref{GS1}) to Eq.~(\ref{GS2}) are valid unless there is a strong mass degeneracy between $N_1$ and one of the scalars. On comparing Eqs.~(\ref{GS1})-(\ref{hdecay}) with Eqs.(\ref{wdecay}), and (\ref{zdecay}), it is observed that the W and Z decay widths are quite large as compared to the scalar decay widths. Also, the tininess of the scalar decay widths could not be redeemed even if one goes on to vary the scalar mass $(200\, \text{GeV}- 1\, \text{TeV})$ the decay widths remain small. In fact, the sum total of decay widths of all the scalars combined is tiny in comparison to the individual decay widths of either Z or W boson. Hence, the relic abundance will solely depend on SM gauge boson decays while the scalar contributions will remain negligible. This is another novel observation of this study and the resulting phenomenological implications are discussed ahead. It should be pointed out that the annihilations producing $N_1$ are significantly suppressed ($\sim \epsilon_{i1}^4$) compared to decays ($\sim \epsilon_{i1}^2$), and hence are excluded from the analysis. Back reactions involving $N_1$ are also not included since initially the $N_1$ number density is vanishingly small. For the same reason, terms proportional to $Y_{N_1}(z)$ are also dropped, which is a standard approximation for the freeze-in case~\cite{Merle2014,Hall:2010}.

The freeze-in production of dark matter ($N_{1}$) in this scenario has been studied quantitatively by numerically solving the full Boltzmann Eq.~(\ref{1BE}). In the Appendix the analytical solution has been presented to clearly observe the dependence of relic density on model parameters. However, for concreteness, numerical approach has been employed. The analysis begins with the standard assumption of the freeze-in scenario that, at high temperatures, the co-moving number density ($Y_{N_{1}}$) of dark matter is zero. Therefore, the  Eq.~(\ref{1BE}) is solved with the initial condition $Y_{N_1}(z=0)=0$ for $T \gg m_{h}$.

\begin{figure}
\begin{center}
\includegraphics[width=0.5\textwidth]{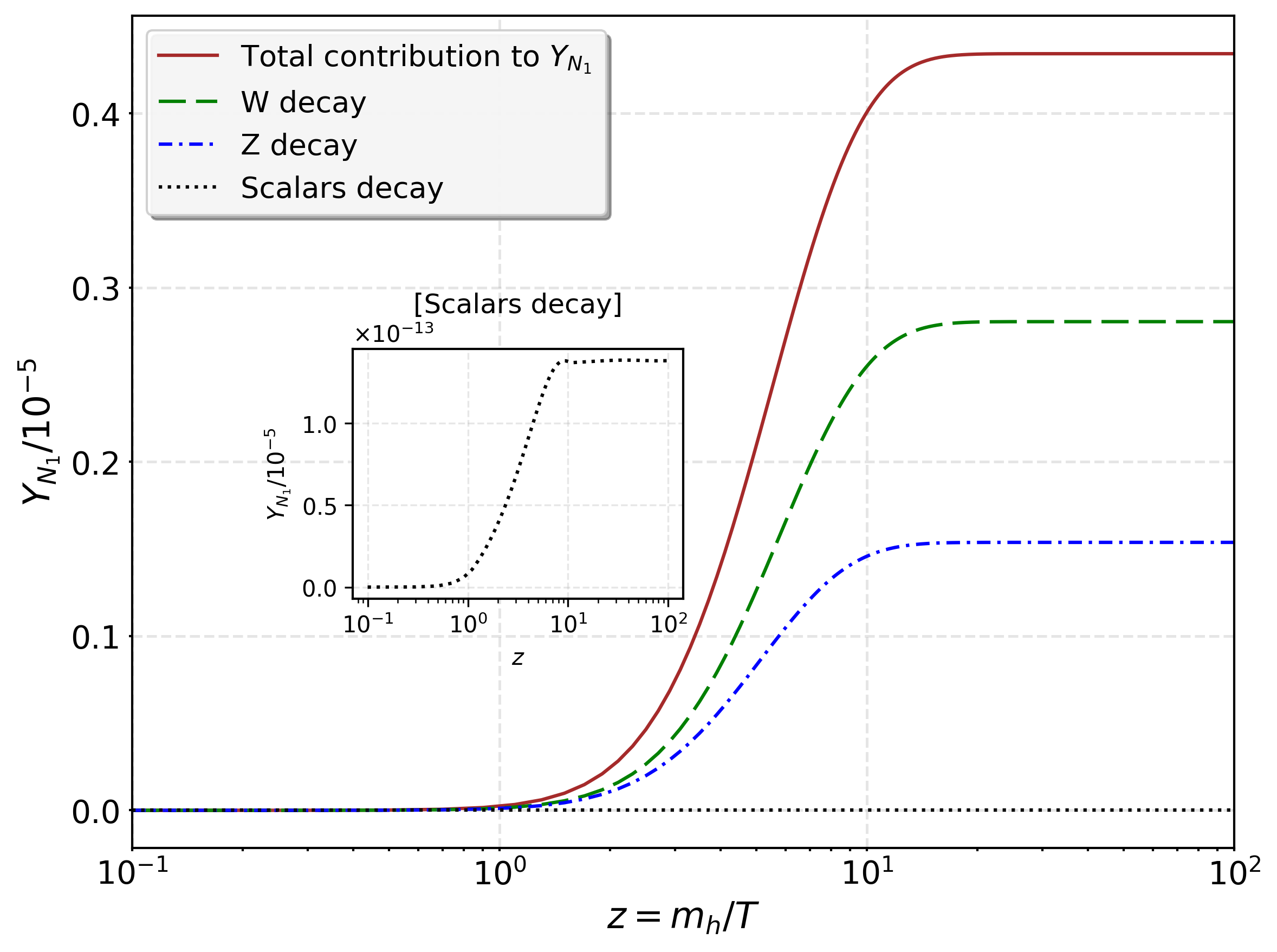} 
\caption{\it Abundance of $N_1$ as a function of temperature from different gauge boson decays. The solid red line represents the total abundance of $N_1$ obtained by summing all contributions at benchmark points $v_{\nu}=10$ MeV, $M_{N_{1}}=0.1$ MeV, and $Y^{\nu}_{i1} \sim 10^{-14}$. The small plot shows the variation of all scalars, which is  suppressed by the factor of $\sim 10^{13}$ from the $W^\pm$ and $Z$ contributions to the total abundance of $N_1$. }
\label{numbden}
\end{center}
\end{figure}

We consider all relevant production pathways of $N_1$ as shown in R.H.S of Eq.~(\ref{1BE}), which consists of two-body decays from all the scalars, $W$, and $Z$ bosons.
In Fig.~\ref{numbden}, the variation of the co-moving number density $(Y_{N_{1}})$ of $N_1$ versus $z$,from specific decay modes and also after considering all the modes, is shown, where $v_{\nu}=10$ MeV, $M_{N_{1}}=0.1$ MeV. 
The extra scalars present in the model are chosen to be degenerate $(m_{A^{0}} \sim m_{S^{\pm}} \sim m_{S^{0}}= 200 \text{GeV})$ for simplicity  and the corresponding Yukawa couplings of $N_1$ are chosen as: $Y^{\nu}_{i1} \sim 10^{-14}$ (this choice is in complete accordance with the non-thermal condition of Eq.~(\ref{ybound})). The solid maroon line corresponds to the full solution of Eq.~(\ref{1BE}) considering all the modes and for the chosen parameters we obtain the correct relic abundance within the $3 \sigma$ of DM relic density ~\cite{Planck2021erratum,Planck2018a}. The dashed green, and blue lines show the individual contributions to $Y_{N_1}$ coming from $W$ and $Z$ boson decays and the horizontal dashed black line at the very bottom shows the combined scalar contributions and have shown in smaller subplot also. It has marginally less contribute (suppressed by a factor of $10^{13}$) to the total $(Y_{N_{1}})$. This clearly depicts the earlier stated result that $N_1$ production is dominated only by the SM gauge boson decays while the scalar contributions remain negligible owing to their small decay widths. Here, we would like to state that increasing the extra scalar masses will not change this result because of their small decay widths. Also, from the nature of the curves, one can observe the standard FIMP behavior: at high temperatures, $N_1$ has a vanishing $Y_{N_{1}}$, but as the Universe expands and cools down (i.e., with increasing $z$), $N_1$ gets slowly produced over time and $Y_{N_{1}}$ gradually increases, with maximum production occurring around $z \sim 2-5$. As the temperature drops further to lower values, $Y_{N_{1}}$ gets \textit{``freezes-in"} and thereby attains a constant value.

In order to compute the relic abundance $\Omega_{N_{1}} h^2$ of the sterile neutrino dark matter one needs to find the value of its co-moving number density $Y_{N_{1}}(z)$ at the present epoch ($T=T_{\infty}$, $T_{\infty}=2.73$K). This value ($Y_{N_{1}}(z=z_{\infty})$) is obtained by solving the Eq.~(\ref{1BE}) for the number density of $N_1$ and noting the epoch at which $Y_{N_1}$ attains a constant value. The expression of $\Omega_{N_{1}} h^2$ in terms of $Y_{N_{1}}(z=z_{\infty})$ is given as~\cite{Edsjo1997}:
(see \ref{appx} for full derivation)
\begin{equation}\label{relicbound}
\Omega_{N_1}h^2 = 2.744\times 10^8 \bigg(\frac{M_{N_1}}{\text{GeV}}\bigg)Y_{N_1}(z_{\infty})
\end{equation}

\begin{figure}[t!]
\begin{center}
\includegraphics[scale=0.48]{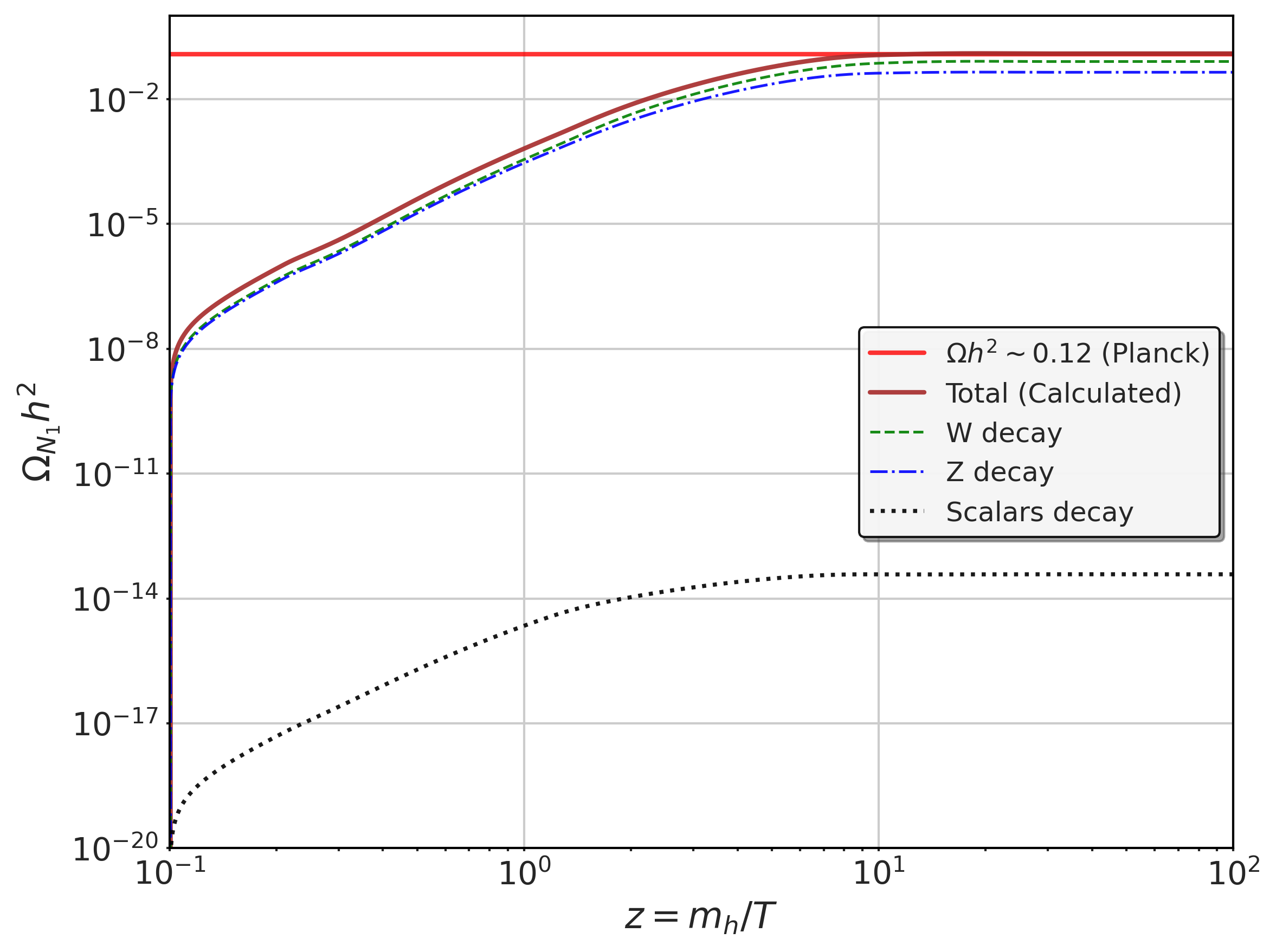} 
\caption{ \it Evolution of the dark matter relic abundance $\Omega h^2$ as a function of $z = m_h/T$. The solid maroon curve denotes the total relic abundance from all decay channels, while the green dashed line shows the contribution from $W$ boson decays, the blue dash-dotted line indicates the $Z$ boson decays, and the dotted black line corresponds to combined scalar decays. The horizontal red solid line indicates the observed relic abundance, $\Omega h^2 \sim 0.12$, as reported by Planck.(benchmark same as Fig.~\ref{numbden}) }
\label{relic}
\end{center}
\end{figure}

In Fig.~\ref{relic}, the variation of relic abundance versus $z$ is presented, illustrating the total as well as relative contributions of gauge bosons and all the scalars for $v_{\nu}=10$ MeV, $M_{N_{1}}=0.1$ MeV. Again, it is clear from the plot that the major contribution comes from the SM gauge bosons, depicted by the dashed green line for the $W$ boson and the dashed blue line for the $Z$ boson, with the $W$ boson contribution slightly leading over the $Z$ boson at any given temperature. The combined scalar contribution is shown by the dashed black line, whose contribution to the relic density at a given temperature is quite low  as mentioned earlier. The total relic density is depicted by the solid maroon line, which sums over all the contributions, and this matches the required relic abundance of dark matter as shown by the horizontal solid red line.

\begin{figure}[t!]
\begin{center}
\includegraphics[width=0.5\textwidth]{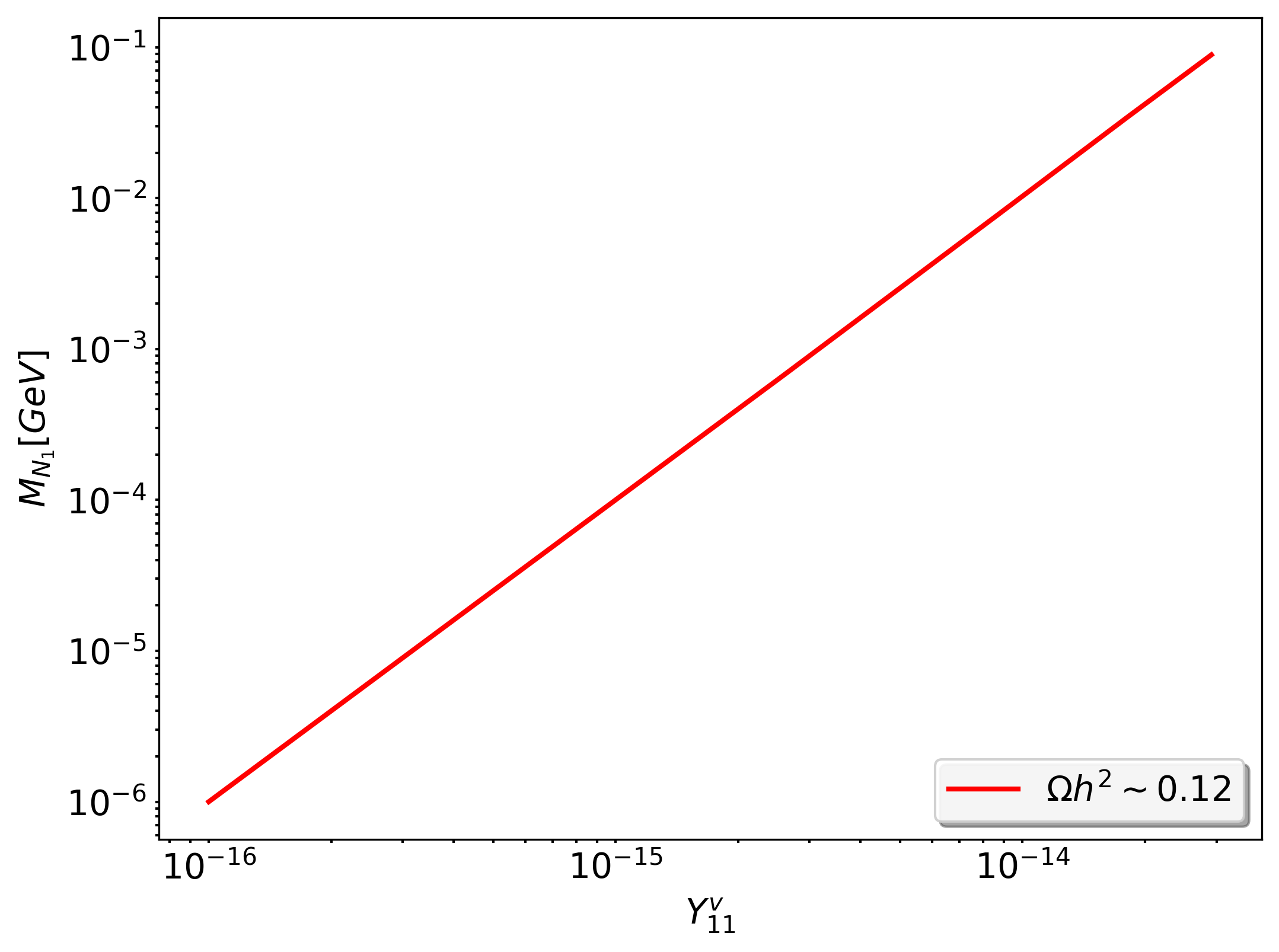} 
\caption{ \it The red solid curve denotes the values of Dirac Yukawa coupling $Y^{\nu}_{11}$ and sterile neutrino mass $M_{N_1}$ for which the observed relic abundance is satisfied via freeze-in production. For simplicity, it was assumed that $Y^{\nu}_{i1} \sim Y^{\nu}_{11}$ where $i$ goes from $1$ to $3$ and benchmark points are $v_\nu \sim 10$ \text{MeV} and $m_{S^\pm} \simeq m_{S^0} \simeq m_{A^0} \sim 200 $ \text{GeV}.}
\label{FigS}
\end{center}
\end{figure}

Hence, despite the presence of some extra heavy scalars in the model coming from linear combinations of the two complex scalar doublets, the major contribution proceeds via gauge boson decays and hence mimics the analysis performed in \cite{Datta:2021elq}. However, this feature was overlooked in previous studies regarding sterile neutrino production in the context of neutrino-philic two-Higgs-doublet model ~\cite{Adulpravitchai:2015mna}. This is because the production via gauge bosons utilizes the heavy-light mixing, which is generally very small as it is suppressed by heavy right-handed neutrinos. However, under the FIMP scenario, the couplings of the dark matter particle are indeed required to be quite small in order to respect the non-thermality condition, so $W$ and $Z$ decays should not be excluded from the analysis. From Eq.~(\ref{relicbound}), one notes that the relic abundance is evaluated to be proportional to the mass of the dark matter candidate $M_{N_{1}}$. However, information about the relevant Yukawa couplings remains hidden in the decays that enter the Boltzmann equation ~(\ref{1BE}). To study the intricate relationship between relic abundance $\Omega_{N_{1}} h^2$, $M_{N_1}$ , and the Yukawa couplings of the first column of $M_D$ ($Y^{\nu}_{i1}$).  We vary the Yukawa couplings $Y^{\nu}_{i1}$ and $M_{N_1}$ simultaneously and enquire about the combinations that give rise to the correct relic abundance, which are shown in Fig.~\ref{FigS}. Generally, one would expect many points spanning an ample region of the parameter space. However, on the contrary, a straight line is found corresponding to those values of ($Y^{\nu}_{i1}, M_{N_1}$) that yield the appropriate relic satisfaction. One also note the rather interesting result that not all values of the Yukawa couplings lead to the correct relic, implying that there is not enough freedom in varying the Yukawa couplings. Nevertheless, there is substantial scope over the choices of $M_{N_{1}}$ values, ranging from keV to a few hundred MeV. This behavior suggests that the relic abundance of dark matter depends heavily on the Yukawa couplings $Y^{\nu}_{i1}$ and is not very insensitive to the mass of the dark matter.

This can be explained by the fact that $Y^{\nu}_{i1}$ is directly responsible for the production of $N_1$. Since, the major production occurs via two-body gauge boson decays, which, based on Eqs.~(\ref{wdecay}) and~(\ref{zdecay}), show that

\[
\Gamma_{W,Z} \propto \frac{|Y^{\nu}_{i1}|^2}{M_{N_1}^2} . 
\]

Using analytical estimates ~\cite{Hall:2010, Adulpravitchai:2015mna} it also follows that 

\[
Y_{N_1}(z_{\infty}) \propto \Gamma_{W,Z} \propto \frac{|Y^{\nu}_{i1}|^2}{M_{N_1}^2}. 
\]

Using Eq.~(\ref{relicbound}), the relic abundance is then determined as: 
\begin{equation*}
    \Omega_{N_1} h^2 \propto \frac{|Y^{\nu}_{i1}|^2}{M_{N_1}},
\end{equation*}
Hence, due to the quadratic dependence on the Yukawa couplings, relic density is more sensitive towards it, and this explains the observed behaviour in Fig.~\ref{FigS}. So, there is not much freedom to vary the Yukawa couplings, while the DM mass ($M_{N_1}$) could assume a wide range of values.

\begin{figure}[t!]
\begin{center}
\includegraphics[width=0.52\textwidth]{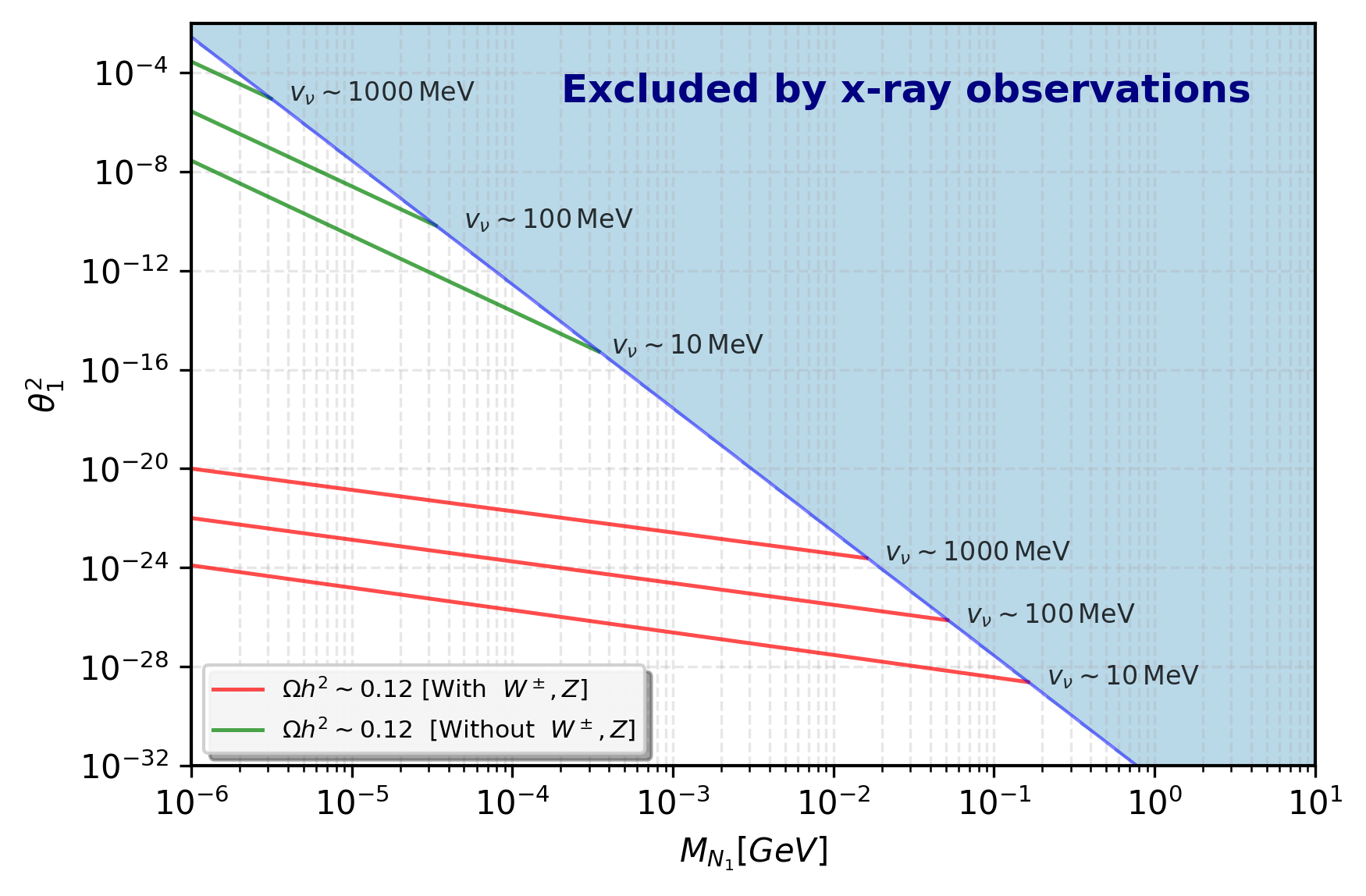} 
\caption{ \it The solid red and green contours denote the regions satisfying the dark matter relic density with and without gauge bosons ($W^\pm, Z$), respectively. The corresponding $v_\nu$ values are indicated at the end of each contour in the $\theta^2_{1}$–$M_{N_1}$ plane. The blue-shaded region is excluded by X-ray observations arising from the decay $N_1 \rightarrow \gamma \nu$  with  $m_{s^+} \simeq m_{S^0} \simeq m_{A^0} \sim 200$ GeV.}
\label{theta}
\end{center}
\end{figure}

\subsection{Parameter Space Enhancement via  VEV ($v_{\nu}$)}
In order for $N_1$ to be a successful dark matter candidate, apart from satisfying relic abundance requirements, it should be stable over cosmological time scales. The best approach is to make the dark matter completely stable, meaning no decay channels should be available. However, in the present scenario, $N_1$ can decay through different channels due to its mixing with light neutrinos via heavy-light mixing. This leads to a decaying dark matter scenario, which generally introduces complications; nevertheless, a distinct advantage is that the decay products could enhance the detectability prospects of the dark matter candidate.

Hence, heavy-light mixing plays a key role in producing the dark matter, destabilizing it, and enhancing the testability of the setup. Regarding the decay products, depending on the mass of $N_1$, the only possibilities are charged leptons (mainly electrons and muons), light neutrinos, and photons. The different decay possibilities, including three-body decays, are given as follows:

(a) via off-shell $W/Z$: 
\[
N_1 \to \ell^-_1 \ell^+_2 \nu_{\ell_2},\, 
N_1 \to \ell^- q_1 \bar{q}_2, \,
N_1 \to \ell^- \ell^+ \nu_\ell, \,
N_1 \to \nu_\ell \bar{\ell'} \ell', \,
\]
\[
N_1 \to \nu_\ell q \bar{q}, \quad 
N_1 \to \nu_\ell \nu_{\ell'} \bar{\nu}_{\ell'}, \quad 
N_1 \to \nu_\ell \nu_\ell \bar{\nu}_\ell ;
\]
(b) via off-shell Higgs: 
\[
N_1 \to \nu_\ell \bar{\ell} \ell;
\]  
(c) radiative decay : 
\[
N_1 \to \gamma \nu.
\]  
Ensuring that that the expected lifetime of $N_1$ must be greater than the age of the Universe, these decay channels will lead to strong constraints on the heavy-light mixing. It turns out that the most stringent constraint is obtained from the radiative decay to photon and light neutrino (c), which translates into a bound ~\cite{Boyarsky2009,Barger1995b,Barger1995a,Pal1982} on the active-sterile neutrino mixing $V_{i1}$ as:
\begin{equation}
\theta^2_{1} = \sum_{i=1}^{3} |V_{i1}|^2 \leq 2.8 \times 10^{-18} 
\left( \frac{\text{MeV}}{M_1} \right)^{5}.
\label{eq:theta_bound}
\end{equation}
So, in the subsequent analysis of dark matter, the above equation is used as a strict bound for the combination of $N_1$ mass and the corresponding Yukawa couplings. 

In Fig.~\ref{theta}, the contour plot for the relic abundance (allowed region) in the $\theta_1$–$M_{N_{1}}$ plane is shown with a solid red and green contours, while the excluded region is indicated in blue, based on the above Eq.~(\ref{eq:theta_bound}) for the radiative decay of the sterile neutrino, as inferred from X-ray observations. The red contours indicate the parameter values that yield the required relic abundance when gauge bosons ($W^\pm,Z$) are included, while the green contours show the corresponding values when gauge bosons are excluded, for different values of the VEV. This demonstrates how the allowed parameter space for the $N_1$ mass expands, when gauge boson channels are taken into account. Also, another kind of enhancement in the DM mass ($N_1$) is observed for lower values of VEV ($v_{\nu}$) of the neutrino-philic doublet. This points towards a very interesting scenario that the relic abundance is directly dictated by VEV. Using Eq.(A15) one can write that
\begin{equation*}
    \Omega_{N_1} h^2 \propto \frac{ M_{N_1} \theta_1 ^2}{v_{\nu}^2}
\end{equation*}
This relation clearly shows that an enhancement in $M_{N_1}$ could easily be obtained by reducing $v_{\nu}$. This is clearly illustrated in Fig. \ref{theta} for different choices of $v_{v_\nu}$ values. This plot also depicts that not all values of the pair ($Y^{\nu}_{i1}, M_{N_1}$) are allowed and that they are highly constrained by the stability requirements of $N_1$. Although $N_1$ masses below the keV scale are allowed by the setup, they are excluded due to the Tremaine-Gunn bound ~\cite{Tremaine1979} on sterile neutrino mass. Thus, the lower limit for the $M_{N_{1}}$ mass is taken to be $1$ keV, while the maximum allowed limit could be increased even up to GeV scale by appropriate choice of $v_{\nu}$ and thus clearly depicting an enlarged parameter space for the sterile neutrino mass.

In order to further probe the nature of this dark matter candidate and place it on firm ground, constraints arising from small-scale structure formation must be taken into consideration. This will not only further constrain the model but also provide information on whether $N_1$ constitutes hot, warm, or cold dark matter, and is discussed in the following section.

\section{Constraints from small-scale structure formation}\label{sec:constraint}
For a truly viable dark matter analysis, obtaining the correct relic abundance is generally not sufficient. Dark matter must also allow structure formation ~\cite{Primack1997, Diemand2011}, such as the formation of galaxies. For instance, dark matter should not be too hot as it would disturb cosmological perturbations and inhibit structure formation in the early Universe. However, in the standard picture, a keV-scale dark matter candidate is motivated to be a warm dark matter candidate. 

Despite this, simply having a mass in the keV range and satisfying relic density requirements does not guarantee warm dark matter. This is because the capacity to form small structures depends on two major factors: 

(i) The mass of the dark matter candidate.

(ii) The momentum distribution function of the dark matter candidate. 

The distribution functions depend on the production mechanism and the details of cosmological evolution. Thus, to determine whether the assumed candidate falls under the category of warm, hot, or cold dark matter, one needs to precisely find the corresponding distribution function and constrain it with astrophysical data ~\cite{Murgia2017}. However, this is a highly non-trivial task, and various approximations are generally employed. 

As an approximate measure, a parameter called the free-streaming length $r_{\text{fs}}$ is introduced ~\cite{Boyarsky2009}, through which dark matter is categorized as hot, warm, or cold. In Physical terms, $r_{\text{fs}}$ represents the average collision-free distance traveled by a dark matter particle after production, before gravitational clustering effects become significant. To calculate $r_{\text{fs}}$ , one requires the distribution function of the dark matter. 

The precise form of the distribution was calculated in ~\cite{Adulpravitchai:2015mna} for the neutrino-philic two Higgs doublet model, and the corresponding relationship between the free-streaming horizon and mass is given by:

\begin{equation}
r_{\text{fs}} = 0.047 \text{ Mpc} \times \left(\frac{10 \text{ keV}}{M_{N_1}}\right).
\label{eq:jhepformula}
\end{equation}

\begin{figure}[t!]
\begin{center}
\includegraphics[width=0.5\textwidth]{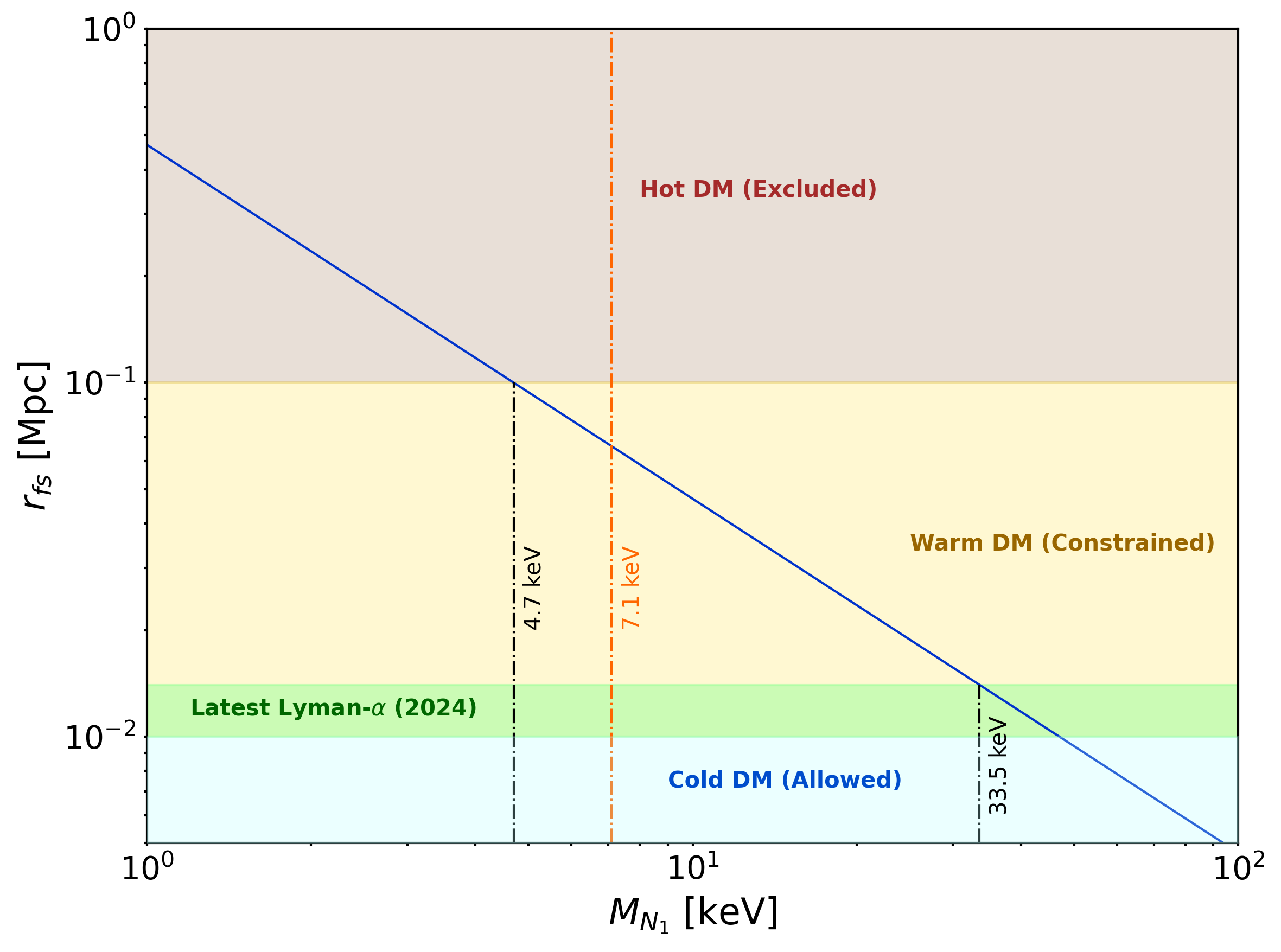} 
\caption{\it The blue line (contour) shows free-streaming length ($r_{\mathrm{fs}}$) as a function of sterile neutrino dark matter mass ($M_{N_1}$), showing excluded hot dark matter (shaded brown), constrained warm dark matter (shaded yellow), further constrained by the latest Lyman-$\alpha$ forest bounds (2024) (shaded green) and allowed cold dark matter (shaded blue) regions. The vertical dashed line marks the $7.1~\mathrm{keV}$ scale relevant to the observed $3.55~\mathrm{keV}$ X-ray line. The plot classifies dark matter candidates by free-streaming horizon consistent with astrophysical constraints.(benchmark same as Fig.~\ref{numbden}) }
\label{regions}
\end{center}
\end{figure}

This is obtained by considering only the scalar decays and assuming degenerate masses for the extra scalars. However, in this study, additional decay processes arise from the interactions involving the W and Z bosons, so, in accurate terms, the distribution function and hence the treatment of the free streaming length should change. However, considering new decay modes for $N_1$ production only leads to the addition of extra distribution functions corresponding to these new modes. Then, the total distribution function of $N_1$ is just the sum of the various distribution functions calculated for different decay modes, and only the coefficients of the distribution functions will change while the polylogarithmic structure will remain intact. Moreover, the phase-space structures of the scalar, $W$, and $Z$ boson decay modes do not differ much as they produce similar daughter particles. This similarity is crucial because if the phase space structures of the various decay processes were diverse, the distribution function could have distinct maxima, and it would become difficult to define mean values, rendering the calculation of the free-streaming horizon futile. For a detailed discussion on these issues, the interested reader is referred to the discussions in \cite{Heeck2017, Boulebnane2018}. 

The classification of sterile neutrino dark matter via the free-streaming horizon is shown in Fig.~\ref{regions}, along with the latest bounds coming from Lyman-$\alpha$ forest measurements \cite{Irsic2024}. The following points are noted:

\begin{itemize}
    \item As noted from Fig.~\ref{theta}, the model predicts a light dark matter mass $\sim 1$ keV, while from Eq.~(\ref{eq:jhepformula}), one obtains $r_{\text{fs}} = 0.4$ Mpc, corresponding to the hot dark matter regime, which is excluded by all observations. For hot dark matter, one must have  $r_{\text{fs}} > 0.1$ Mpc, which corresponds to a mass $M_{N_1} < 4.7$ keV. This condition is not satisfied by  particle acting as a hot dark matter candidate. Hence, the brown-shaded region in Fig.~\ref{regions} is excluded.
   
    \item The blue line (contour) ($4.7 < M_{N_1} < 33.5$) keV, which lies between the corresponding black dashed lines (corresponds mass of $N_1$ $4.7\, \text{keV and} \, 33.5\, \text{keV} $ respectively) and the yellow-shaded region, also produces the desired relic abundance for warm dark matter (see Fig.~\ref{theta}). Using Eq.~(\ref{eq:jhepformula}), this interval corresponds to ($0.014 < r_{\text{fs}} < 0.1$) Mpc, which characterizes warm dark matter. Thus, the model accommodates a warm dark matter candidate if $M_{N_1}$ lies within this range. This region depicts the allowed interval for warm dark matter. However, recent Lyman-$\alpha$ forest measurements further constrain this region, with the latest bound shown by the green-shaded region providing an updated lower limit on the warm dark matter mass. An interesting feature of the setup is its feasibility to explain the claimed X-ray line at $3.55$ keV, since $M_{N_1} = 7.1$ keV is allowed by the relic density and corresponds to warm dark matter. This is clearly shown in Fig.~\ref{regions} by a vertical dashed orange line.
    \item Lastly, for the case $r_{\text{fs}} < 0.014$ Mpc, Eq.~(\ref{eq:jhepformula}) implies $M_{N_1} > 33.6$ keV. This regime, up to $0.2$ GeV, can explain all the observed dark matter in the Universe and corresponds to cold dark matter. This region is shown in Fig.~\ref{regions} as the blue-shaded region and is allowed by all observational data.
\end{itemize}

So, as one goes to a higher mass scale of $N_1$, it follows from Eq.~(\ref{eq:jhepformula}) that one obtains colder dark matter candidate as a result of decreasing $r_\text{fs}$ (Table-\ref{tab:constraints}). However, a novel feature of this setup is that it is able to accommodate both warm and cold dark matter options and an explanation \cite{Bulbul2014,Boyarsky2014} for the much talked $3.55$ keV X-ray line.

\begin{table}[tb]
\centering
\caption{Comparison of structure formation constraints with $\nu2$HDM model predictions.}
\label{tab:constraints}
\begin{tabular}{lcc}
\hline\hline
\textbf{Observation} & \textbf{Bound} & \textbf{Bound ($\nu2$HDM)} \\
\hline
Latest Lyman-$\alpha$ limit & $M_{N_1} > 33.6$ keV & $M_{N_1} \gtrsim 34$ keV \\
Free-streaming length & $r_{\text{fs}} < 0.014$ Mpc & $r_{\text{fs}} \lesssim 0.014$ Mpc \\
Dark matter classification & Viable regime & Safe \\
\hline\hline
\end{tabular}
\end{table}

\section{Discussion and  Conclusions}
\label{sec:con}

Considering the neutrino-philic two-Higgs-doublet model, we have shown that one can simultaneously account for the masses of the active neutrinos, demand a low-scale seesaw, and explain the dark matter problem, where exactly one of the singlet fermions, $N_1$, can be out of equilibrium in the early Universe and act as a FIMP.

We discussed important phenomenological implications of this model that were overlooked in previous studies, especially the dominant role of SM gauge bosons in dark matter production through heavy-light mixing ($V_{ij} = \frac{M_{D_{ij}}}{M_{N_{j}}}$), which leads to interactions between the heavy right-handed neutrinos with the $W$ and $Z$ bosons. In the context of the FIMP scenario, however, the dark matter couplings are inherently required to be extremely small to remain out-of-equilibrium. So, the out-of-equilibrium condition is the holy grail of the freeze-in mechanism. We found that if one incorporates the SM gauge boson decays it leads to a new revised non-thermal condition which predicts further lower values for the Yukawa couplings ($Y^{\nu}_{i1}$) of $N_1$ and leads to a new parameter space. 

Incorporating these issues, we studied the evolution of the comoving number density of $N_1$ as the Universe expands and demonstrated the dominance of $W$ and $Z$ boson decays over scalar decays around order of $10^{13}$. The same feature was observed in the evolution of the relic abundance, with the $W$ boson contribution slightly leading over the $Z$ boson, while the scalar contribution remained negligible. The respective values of the pair ($Y^{\nu}_{i1}$, $M_{N_1}$) were scrutinized by the relic requirements, and it was concluded that the relic density is almost independent of the mass of the dark matter while the Yukawa couplings are tightly constrained. 
The model predicts a wide range for the $N_1$ mass, spanning from $1$ keV to upto GeV scale, for different VEVs ($v_{\nu}$) of the extra Higgs doublet (Fig.~\ref{theta}). This analysis features a novel possibility to obtain a widened parameter space (dictated by small $v_{\nu}$) which can account for all the observed dark matter in the Universe. Also, a particular choice of $v_{\nu}$ fixes the scale of the lightest active neutrino's mass ($m_{\nu_1}$), as from Eq.~(\ref{a15}) one obtains: $\Omega_{N_1} h^2 \sim \frac{m_{\nu_1}}{v_{\nu}^2}$. Hence, the setup leads to a correlation between the relic abundance, low-scale seesaw and mass-scale of the lightest active neutrino ($m_{\nu_1}$). 

Additionally, according to the latest Lyman-$\alpha$ results, the model accommodates both warm and cold dark matter options and provides a plausible explanation for the debated $3.55$ keV X-ray line. We point out that for the mass range $(1-10)$\,MeV, the sterile neutrino will undergo three-body decays via off-shell $W$ and $Z$ bosons, producing an electron-positron pair and a light neutrino. The annihilation of the electron-positron pair could offer a feasible explanation for the observation of the $511$ keV line seen by INTEGRAL/SPI \cite{Siegert:2015knp, Knodlseder:2003sv}.
Other pathways towards testability stem from the prediction of a very light scale for one of the active neutrinos, which could falsify the setup through ongoing (or future) experiments such as KATRIN~\cite{Aker:2021gmm} and PROJECT-8 Collaborations ~\cite{AshtariEsfahani:2017ryh}.

\section*{Acknowledgments}
We sincerely thank Prof. Rathin Adhikari for useful discussions and Ang Liu for initial help in computations. S.K. also gratefully acknowledges Prof. Sushant Ghosh for providing the necessary computational facility used in this work. We also thank the anonymous referee for their thoughtful and constructive comments, which have helped us to improve the clarity and robustness of our manuscript.

\appendix
\section{Analytical Solution of Boltzmann Equation}\label{appx}
 Since, there is only a single differential equation to be solved, we follow the procedure outlined in \cite{Hall:2010}. So, we define the equilibrium number density of a particle `A' as:
\begin{equation}
n_A^{\rm eq}(T) = \frac{g_A}{2\pi^2}\, m_A^2\,T\, K_2\!\left(\frac{m_A}{T}\right),
\end{equation}
where $m_A$ is the mass of the particle and $g_A$ is its degree of freedom. The entropy density is given by:
\begin{equation}
s(T)=\frac{2\pi^2}{45}\, g_s(T)\, T^3.
\end{equation}
Thus, we have the equilibrium yield given as:
\begin{equation}
Y_A^{\rm eq}(T)=\frac{n_A^{\rm eq}}{s}
= \frac{45\, g_A}{4\pi^4\, g_s(T)}
\left(\frac{m_A}{T}\right)^2
K_2\!\left(\frac{m_A}{T}\right).
\end{equation}
Now, we define the following parameters:
\begin{equation}
z=\frac{m_h}{T}, \qquad 
\xi=\frac{m_A}{T} = r z,\qquad r=\frac{m_A}{m_h}.
\end{equation}. We can write the Eq.~(\ref{1BE}) in a simple way as:
\begin{align}
\frac{dY_{N_1}(z)}{dz} &= 
\frac{2 M_{\rm Pl}\, z}{1.66\, m_h^2} 
\frac{\sqrt{g_*(z)}}{g_s(z)}
\sum_A Y_A^{\rm eq}(T)\,
\big\langle\Gamma_{A\to N_1 X}\big\rangle,
\label{eq:BE}
\end{align} where `X' could be a charged lepton ($l^{\pm}$) or an active neutrino ($\nu$) and every species `$A$' represents a decay channel: 
\begin{equation*}
    A = \{ Z,\; W^\pm,\; h,\; A^0,\; S^0,\; S^\pm \}.
\end{equation*}
Keeping one decay channel $A$, the Boltzmann equation becomes
\begin{equation}
\frac{dY_{N_1}}{dz}
= A'\, r^2\, z^3\, K_1(r z),
\end{equation}
where we have defined,
\begin{equation}
A' = 
\frac{2M_{\rm Pl}}{1.66\, m_h^2}
\frac{45\, g_A\, \Gamma_{A\to N_1 X}\, \sqrt{g_*}}
     {4\pi^4\, g_s^2}.
\end{equation}
Now, integrating Eq.(A6) from $z=0$ to $z=\infty$ and using the initial condition $Y_{N_1}(0)=0$):
\begin{align}
Y_{N_1}(\infty) &= A' r^2 \int_0^\infty z^3 K_1(rz)\, dz.
\end{align}
Making a variable change $t=r z$, one gets: 
\[
r^2\int_0^\infty z^3 K_1(rz)\,dz
= \frac{1}{r^2} \int_0^\infty t^3 K_1(t)\, dt.
\]
Utilizing the standard Bessel integral,
\[
\int_0^\infty t^3 K_1(t)\, dt = \frac{3\pi}{2},
\]
this leads to the following expressions for the freeze-in yield:
\begin{equation}
Y_{N_1}^{(A)}(\infty)
= \frac{135}{4\pi^3}\;
\frac{M_{\rm Pl}\, \sqrt{g_*}}{1.66}\;
\frac{g_A\,\Gamma_{A\to N_1 X}}{m_A^2 g_s^2}.
\label{eq:singleYield}
\end{equation} 
above expression applies to both scalars and vectors (only $g_A$ differs). Now, one can sum over all the contributions as:
\begin{equation}
Y_{N_1}^{\rm (tot)}(\infty)
= \frac{135}{4\pi^3}\;
\frac{M_{\rm Pl}\, \sqrt{g_*}}{1.66\, g_s^2}
\sum_A \frac{g_A\, \Gamma_{A\to N_1 X}}{m_A^2}.
\label{eq:totalYield}
\end{equation}
Typical internal degrees of freedom:
\[
g_A =
\begin{cases}
3, & A = Z,\; W^\pm \\
1, & A = h,\; A^0,\; S^0,\; S^\pm.
\end{cases}
\]
Thus the full solution is
\begin{align}
Y_{N_1}^{\infty} &=
\frac{135}{4\pi^3}
\frac{M_{\rm Pl}\sqrt{g_*}}{1.66\, g_s^2}
\Bigg[
\frac{3\,\Gamma_Z}{m_Z^2}
+\frac{6\,\Gamma_W}{m_W^2}
+\frac{\Gamma_h}{m_h^2}
+\frac{\Gamma_{A^0}}{m_{A^0}^2}
\nonumber \\[3pt]
&\qquad\quad
+\frac{\Gamma_{S^0}}{m_{S^0}^2}
+\frac{2\,\Gamma_{S^\pm}}{m_{S^\pm}^2}
\Bigg].
\end{align}
On comparing the decay widths of all the scalars, W and Z boson (see Eq.(\ref{GS1})-Eq.(\ref{zdecay})) it turns out that $W$ and $Z$ decay width are considerably large. Hence, one can safely ignore the scalar contributions and write:
\begin{align}\label{fir}
Y_{N_1}^{\infty} & \simeq
\frac{405}{96 \pi^3}
\frac{M_{\rm Pl}\sqrt{g_*}}{1.66\, g_s^2} (M_{Z}+M_{W}) \frac{ |Y_{i1}^{\nu}|^2}{M_{N_1}^2}
\end{align}

In order to calculate relic abundance we note that the present entropy density $s_0$ and critical density are \cite{ParticleDataGroup:2024cfk}:
\[
s_0 = 2891.2~{\rm cm^{-3}},\qquad
\frac{\rho_{c,0}}{h^2}
= 1.053672\times 10^{-5}~{\rm GeV\, cm^{-3}}.
\]
Thus the relic abundance is
\begin{equation}
\Omega_{N_1}h^2
= 
\frac{m_{N_1} s_0}{\rho_{c,0}/h^2}\;
Y_{N_1}(\infty).
\end{equation}
Evaluating the constants gives
\begin{equation}
\boxed{
\Omega_{N_1} h^2 
= 2.74393\times 10^8\;
\left( \frac{m_{N_1}}{\rm GeV} \right)
Y_{N_1}(\infty)
}.
\label{eq:OmegaBasic}
\end{equation}
From Eq.~(\ref{fir}) we have:
\begin{align}\label{a15}
\Omega_{N_1} h^2 &\simeq
2.74393\times 10^8
\left( \frac{M_{Z}+M_W}{\rm GeV} \right)
\left(\frac{405}{96 \pi^3}\right) \left( \frac{M_{Pl} \sqrt{g_{*}}}{1.66 \, g_{s}^2} \right) \frac{|Y^{\nu}_{i1}|^2}{M_{N_1}}
\end{align}
This shows that the relic abundance is controlled by Yukawa couplings of $N_1$ and its mass.

\bibliographystyle{apsrev4-1}
\bibliography{filmpscot2.bib}

\end{document}